\documentclass[preprint2]{aastex}

\newcommand{\heii}{He~II~$\lambda$4686~\AA}
\newcommand{\etac}{$\eta$ Carinae}
\newcommand{\etas}{$\eta$ Car}

\slugcomment{Accepted in ApJL}

\shorttitle{\heii~in $\eta$ Carinae}
\shortauthors{Steiner \& Damineli}

\begin{document}


\title{ Detection of \heii~in \etac}

\author{J. E. Steiner and A. Damineli\altaffilmark{1}}
\affil{IAG - University of S\~ao Paulo, R. do Mat\~ao 1226, 05508-900 S\~ao
Paulo, Brazil}
\email{steiner or damineli@astro.iag.usp.br}

\altaffiltext{1}{Based on observations collected at the Pico dos 
Dias Observatory (LNA/Brazil)}

\begin{abstract}

We report the detection of the emission line \heii~in \etac.
The equivalent width of this line is $\sim$100 m\AA~along most of the 
 5.5--yr cycle and jumps to $\sim$900 m\AA~just before phase 1.0, 
 followed  by a brief disappearance. The similarity 
 between the intensity variations of this line and 
 of the X--ray light curve is remarkable, suggesting that they 
 are physically connected. We show
 that the number of ionizing photons in the ultraviolet 
 and soft X--rays, expected to be emitted in the shock wave from 
 the colliding winds, is of the order of magnitude required to 
 produce the He~II emission via photoionization.
      
The emission is clearly blueshifted when the line is strong. 
The radial velocity of the line is generally --100 Km~s$^{-1}$, 
decreases steadily just before the event,  and reaches 
--400 Km~s$^{-1}$ at ph$=$1.001. At this point, 
the velocity gradient suddenly changes sign, at the same 
time that the emission intensity drops to nearly zero. 
Possible scenarios for explaining this emission are 
briefly discussed. The timing of the peak of He~II intensity 
is likely to be associated to the periastron and may be a 
reliable fiduciary mark, important for constraining the orbital parameters.

\end{abstract}

\keywords{stars: early-type--stars: individual (\object{\etac})--binaries:general}

\section{Introduction}
Massive stars have strong impact on galactic environments. Their evolution, however,
is not very well known, since their intrinsic parameters are difficult to determine. 
This is even more true for the mass, that requires a binary companion to be weighted. 
The Luminous Blue Variable (LBV) \etac, believed to be one of the most massive stars 
known from its luminosity, has shown signatures of binarity, opening the possibility 
to measure this fundamental parameter. 

The first evidence of binarity emerged from the true periodicity in the 5.5 year
cycle \citep[in preparation]{D96,v03,W04,MC04}. 
The second step to unveil the binary nature of the star was done when specific models
could be calculated. Damineli, Conti \& Lopes (1997, hereafter DCL) derived highly eccentric orbits 
from radial velocities of Paschen lines and suggested that X--rays are produced 
by wind--wind collision. That particular model had problems, since emission lines 
do not trace precisely the orbital motion \citep{D97}. However, parameters 
similar to that of DCL were used to reproduce the main features of the X--ray emission 
in \etas~by wind--wind collision models \citep[hereafter PC]{MC01,PC02}. 
The orbital parameters, however, are still not well constrained.

The subject of this Letter is to present the unexpected detection of 
the \heii~variable emission line in \etas~and to explore how it is related 
to the binary nature of the star. 

\section{Observations and Results}

The data presented here are part of a  long--term spectroscopic monitoring of \etas,   
started in 1989 at the Coud\'e focus of the 1.6--m telescope of the Pico dos Dias 
Observatory (LNA/Brazil). Full results, including the analysis of many low and high 
excitation lines, will be reported in Damineli et al. (2004, in preparation), 
from which we are using the period length, P$=$2022.1--d (5.536--yr) and the time 
of phase zero (ph$=$0.0) of the spectroscopic event (JD$=$2452819.8, 29 June 2003), 
defined by the disappearance of the narrow component in the He~I 6678 \AA~line.

\begin{figure}
\plotone{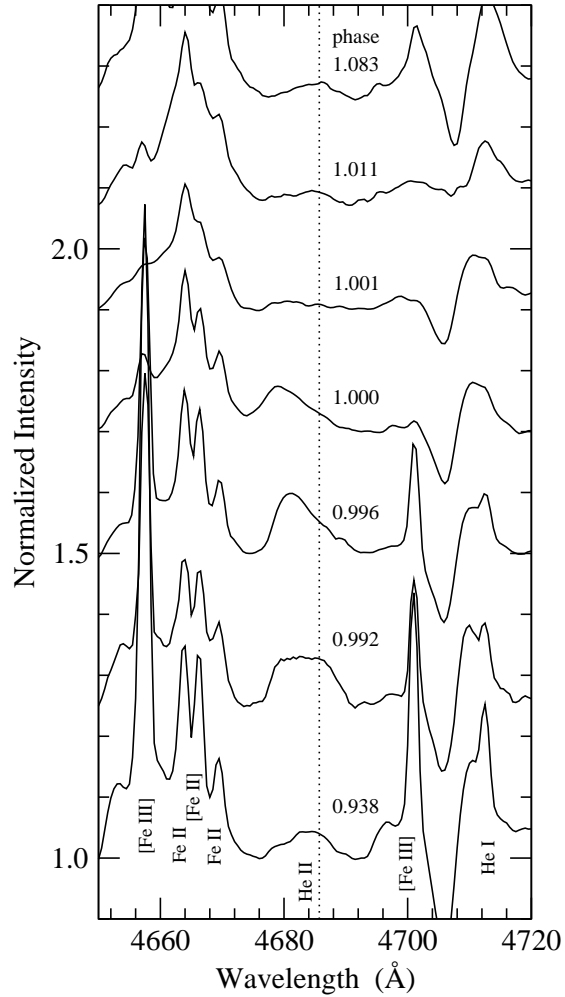}
\caption{Sample of \heii~line profiles labeled by phase of the 2022.1--d period. The vertical dotted line indicates the rest
wavelength of \heii. \label{fig1}}
\end{figure}

 Data reduction was done with IRAF package in the standard way. The spectral 
 resolution was degraded to 0.5\AA/pixel in order to enhance the 
 signal--to--noise ratio. Most of the spectra used here have S/N$>$100 in the 
 stellar continuum and some of them twice as much, specially around 
 ph$=$0.2--0.8, when many spectra were co--added. We detected an emission 
 line at $\sim$4680--85 \AA~(Figure 1), that we suggest to be 
 \heii.  We searched for possible transitions 
 in the range 4670--4700 \AA~and found none from species that are typical of 
 the \etac~spectrum.  \heii~has never been detected with certainty 
 before, as discussed by Hillier and Allen (1992 
 and references therein), who reported an upper limit of 1\AA.  
 The line showed up in almost all of our high quality spectra, 
  except in a few ones of lower S/N. Paradoxically, 
 this line is faint in high excitation phases 
 and strong during the event, contrarily to the behavior of the high excitation 
 forbidden lines. In Figure 1 we display a sample of spectra collected in 2003, 
 ordered by phase of the 2022.1--d period.
 
 In order to evaluate the errors, we measured several times each line, with different 
 assumptions for the continuum. The errors were 20--100 m\AA~in the equivalent widths 
 (EWs)  and 10--110 Km~s$^{-1}$ in the radial velocities (RVs) of the line centroid 
 (V$_{cen}$) and are displayed in Figures 2 and 3. Those are formal errors; 
 systematic errors may also be present, for example 
 introduced by the rectification of the  continuum.  The fit to the continuum 
 was done by a 3rd order polynomial, what introduces only low frequencies 
 and has little influence on scales comparable to the extent of a  spectral line. 
 Structures in the flatfielding image could introduce distortions in the continuum,
 although of low frequency, since we used always the same observational
 setup and data reduction procedures. The EWs are affected by an additional type of
 systematic error: line blending, that prevents the assessment of the local continuum 
 and consequently the extension of the line wings. This makes the measurements of EWs
 and FWHM ($\sim$500--600 Km~s$^{-1}$) somewhat underestimated, as is the case for 
 almost all other spectral lines in the rich spectrum of \etas.
 The position of the  line centroid is much more robust and is 
 affected only by the S/N ratio, consequently
 the errors can be better judged from the overall scatter in the RV curve.
 
 The slit width was kept fixed at $\sim$1.5", and this seems 
 relevant, since a fraction of the He~II emission appears to be extended. 
 Our data of 1997 and 2002, when compared with  contemporaneous 
 spectra taken with FEROS spectrograph at ESO ($\sim$3.6" fiber 
 entrance), result in smaller EWs. The comparison of our 2003 data with 
 those taken with STIS on board of HST ($\approx$0.1" slit width) result 
 in larger EWs, indicating that the EWs are larger for wider slits. 
 This is the opposite of that reported by Hillier and Allen (1992) for
  H~I, He~I and Fe~II lines in the Homunculus, which are smaller than 
  those in the central object. He~II seems to be intrinsically in emission 
  near the central source, since dust scattering would preserve the equivalent widths. 
  The line shape and radial velocity are unchanged with the slit aperture,
 indicating that most of the emission should arise from the central object.
 In fact, long--slit CCD HST/STIS spectra (52"x 0.1") collected prior 
 to the 2003.5 event (T.R. Gull, private communication) show that the He~II
 emission was confined to 0.1" (2--pixel limit) in the East--West direction.
 However, the emission could be extended in other directions outside 
 the slit. Although this question is relevant, the data available to us 
 are insufficient to derive any firm conclusion and we will 
 restrict our analysis to the homogeneous Brazilian set of data.
  
In Figure 2 we see that the strength of the He~II line is very weak 
 along the 5.5--yr cycle at the level of EW$\approx$100 m\AA. It starts 
 strengthening around ph$\approx$0.8, jumps to EW$=$873 m\AA~just
  before ph$=$1.0, and after a brief fading, it rises again, reaching 
  a local maximum (EW$=$300 m\AA) 
at ph$\approx$1.04. In  Figure 3a we display a zoom in the EW curve 
around ph$=$1.0, showing a peak centered at ph$=$0.994, a fast drop after
 ph$=$1.0 and a minimum around ph$=$1.003. In Figure 3b we show the RV curve
as measured from the velocity of the line centroid. The velocity becomes 
more and more negative as the system approaches ph$=$1.0, reaching a sharp minimum 
(V$_{cen}$=$-$396 Km~s$^{-1}$) at ph$=$1.001. After ph$=$1.002, uncertainties 
in RVs increase significantly and we will not comment them any further.

\section{Interpretation and Discussion}

	The interpretation of the intensity curve of \heii~as well as of its 
radial velocity may provide important insights about the nature and structure of the 
system. The first thing to notice is the strong similarity between the He~II emission 
curve and the X--ray light curve \citep{MC01,I99}. Both 
emissions start to enhance at ph$=$0.8, peak just before ph$=$1.0, drop to near zero 
intensity at about ph$=$1.001 and return to normal intensity after ph$\sim$1.1. A 
noticeable difference between the two light curves is that He~II has a much 
narrower and prominent peak (centered at ph$=$0.994) than the X--ray light curve, 
which has its maximum at ph$=$0.985. The X--ray emission is thought to be 
originated from the shock wave of the colliding winds in a binary system. Its 
"eclipse" has been modeled by PC, who 
derived system parameters that will be adopted as a reference frame in the 
following discussion. Given this similarity, it is tempting to interpret the He~II 
emission as being originated from photoionization by X--ray or by the 
ultraviolet (UV) photons associated to this source.

\begin{figure}
\plotone{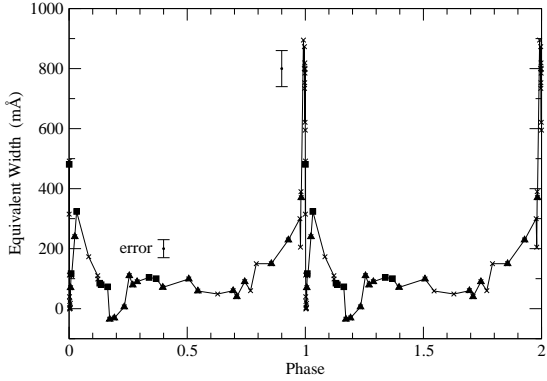}
\caption{\heii~line EW curve for the last three cycles - $\blacktriangle$s are for 1997/98;
   $\blacksquare$s for 1992; and {\bf{x}s} for 2003. \label{fig2}}
\end{figure}

\begin{figure}
\plotone{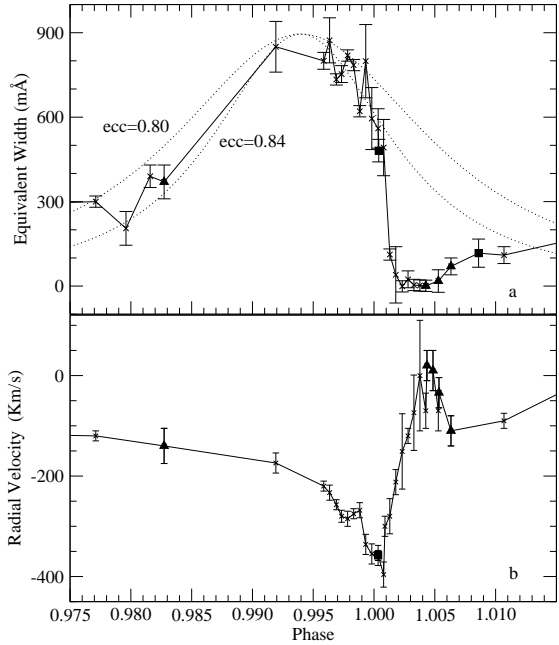}
\caption{Zoom in \heii~variations around ph$=$0. Upper panel: {\it solid line} 
EW curve; {\it dotted lines} He~II emission models. Lower panel: RV curve. 
Same symbols as in Figure 2 \label{fig3}}.
\end{figure}

 Near periastron, the X--ray and UV ionizing photons are mostly unobserved 
as they are absorbed by the intervening gas along the line of sight. 
However, gas near the shock wave (including the winds of the primary and of the 
secondary stars)  may still be directly exposed to this ionization. The \heii, at its 
maximum intensity, has a luminosity of $L_{HeII}$$\sim$100 L$_{\odot}$ 
(assuming the somewhat uncertain values of V$\sim$7.5 and Av$\sim$6 for the 
central source). This amounts to about 9$\times$10$^{46}$ photons per second. 
In order to produce this emission, a luminous 
source of ionizing soft X--ray and ultraviolet photons is required; the peak luminosity 
of the unabsorbed X--rays in the range 2--10 kev is only L$_{Xunabs}$$=$67 L$_{\odot}$ 
\cite{I99}. Detailed hydrodynamical numerical calculations by PC show that 
when the velocity of the secondary's wind is large (V$_{2}$$\sim$3000 Km~s$^{-1}$) 
the energy spectrum behaves like a power law, at soft X--rays. From their Figure 3 we 
 derive a spectrum of L$_{\epsilon}$$=$2.5$\times$10$^{35}\epsilon$$^{-2.7}$ 
 ergs~s$^{-1}$~kev$^{-1}$  for the system parameters favored by those authors 
($\dot M_{2}$$=$10$^{-5}$ M$_{\odot}$ yr$^{-1}$  and V$_{2}$$=$3000 Km~s$^{-1}$). 
In the case that this spectrum can be 
extrapolated down to 54 eV (He$^{+}$ threshold ionization), the total luminosity would 
be L$_{ion}$$=$5200 L$_{\odot}$. This extrapolation seems to be reasonable since the energy loss by 
the wind from the secondary star alone is $\dot E$$_{2}$$=$7500 L$_{\odot}$. The total number 
of photons emitted by this source is N$_{ion}$$=$2.5$\times$10$^{47}$
 ph~s$^{-1}$ what amounts to $\approx$3 times the observed number of \heii~photons. 
 If we add to this the energy from the shock of the primary star, this number will be 
 even larger. We conclude that near periastron the production of ionizing photons is 
 of the right order of magnitude to produce the observed He~II emission 
 via photoioniozation. It is worth noting that only $\approx$1\% of the kinetic 
 energy from the wind of the 
 secondary star is radiated in the 2--10 kev band. Where are the other 99\% going to? 
  The detection of He~II emission may be the first hint that they are mostly radiated 
  in the UV.

The  fact that the bulk of the He~II emission is short lived in terms of the EW 
curve (lasting for $\approx$1\% of the orbital cycle) suggests that its peak indicates the 
passage of the periastron -- and this is quite interesting. This timing is not precisely 
determined either by the X--ray light curve (as the wind of the primary star 
is optically thick to X--ray absorption) or by the high excitation
 emission lines (that are emitted far away), although the low ionization 
 event is generally believed to be associated to the periastron 
approach.  The timing of the periastron (ph$=$0.994) would be the second orbital 
parameter determined with accuracy, besides the period.

A solution to the X--ray light curve was obtained by PC in which the wind--wind shock 
before eclipse is observed through the opening angle in such a way that the X--ray 
source is seen through  the non-absorbing wind from the secondary star. After the
 shock front passes through the line of sight, the X--rays are absorbed by the wind 
 from the primary star. In this way the symmetry of the observed emission is broken
  and one gets a post--minimum flux that is lower then the pre--minimum one. If we 
  assume that the periastron passage occurs at ph$=$0.994, we are able to fit the 
  X--ray light curve with the parameters ecc$=$0.84 and the longitude of the periastron 
  of the primary star, $\omega$$=$212$^{o}$. These parameters predict a superior 
conjunction for the secondary star with a true anomaly of $\theta$$=$58$^{o}$  at 
the phase where our observations show a sharp drop in EWs and a gradient reversal 
in the RV  curve.  Does this suggest an eclipse? This question brings us to another 
one: where is the region of  He~II emission located? 

The large values of the velocity 
 cannot be associated to the shock wave itself as the gas in that region has 
 nearly the velocity of the center of mass. A similar argument could be made 
 with respect to the wind of the primary star. The terminal velocity of the
 primary star is about $\sim$500 Km~s$^{-1}$; at the phase under consideration, 
the shock wave is quite near the primary and the velocity of the wind is still far from 
 terminal. In addition, the shock wave is symmetric with respect to the orbital
 plane and the line of sight imposes a projection of i$\sim$45$^{o}$. 
In this scenario, it seems difficult to combine the emission and radial 
  velocity behavior as 
displayed in Figure 3. In particular, the radial velocity is bluest when the emission 
has dropped to its half maximum intensity, after reaching the peak. If we suppose 
that the emission comes from the near side of the primary star, ionized to  He$^{++}$ by 
the secondary or by the shock, one would expect maximum blue-shift at maximum emission.

Perhaps a more promising explanation is that the He~II emission comes from the wind of
 the secondary star, ionized by the external UV emission from the shock wave. This 
 would produce an ionization front that divides this wind into a zone of  He$^{+}$  and 
 a zone of  He$^{++}$. The  He$^{++}$ zone would be directly exposed to the UV emission 
 and an asymmetric configuration is produced with respect to the secondary star. This 
  asymmetry will cause a blue-shifted emission when the system is seen at an 
appropriate angle. As periastron occurs at ph$=$0.994 and the longitude of 
periastron is $\omega$$=$212$^{o}$, the ionized He~II emitting wind from the secondary 
star should naturally display negative velocities when seen from the line of 
sight of the observer. The terminal velocity of the wind from the secondary star is 
about $\sim$3000 Km~s$^{-1}$ (PC), so one might expect higher velocities and line widths 
than observed. However, the emission  is likely to come from deep into the 
wind where the density is higher and the velocity smaller, still far from the 
terminal velocity. In addition, the inclination of the system is about i$\approx$45$^{o}$,
 which reduces the velocity by about $\sim$cos{\it i}. 
 Let us assume a highly idealized model of an ionized hemisphere with concentric 
 layers of radii r and normal velocity law: v$=$v$_{0}$(1-R$_{2}$/r)$^{\beta}$ 
 (where R$_{2}$ is the radius of the secondary star). 
  As the emissivity is proportional to the square of the density, the 
  contribution of each layer to the line intensity is proportional 
  to $\sim$r$^{-2}$. The peak of the line profile should, therefore, 
  be emitted at velocities that correspond to the innermost 
  ionized zone. We estimate that in order to produce the observed RVs 
  (--400 Km~s$^{-1}$), the innermost ionized zone should 
  occur at r$\approx$ 4 R$_{2}$, for $\beta$$=$1.
    At periastron, the distance from the secondary star to the innermost 
    zone is 60 R$_{\odot}$, to the stagnation shock front is 210 R$_{\odot}$ 
    and to the surface of the primary, 360 R$_{\odot}$. 
    
  The observed line widths (FWHM$\sim$500--600 Km~s$^{-1}$) require a radius that is 
  somewhat smaller. However, as pointed out before, the observed
  FWHM is a lower limit because the line wings/continuum are difficult to 
  be determined  with accuracy. If the He~II emitting wind is indeed eclipsed, 
  one would expect that when the intensity is at its half 
  maximum, half of the wind should 
be eclipsed and the velocity should be most negative; this is so because, given 
the structure of the shock wave and its skew angle (see PC), at this time the 
portion of the wind with more positive velocity is already eclipsed. This 
prediction is confirmed as both events occur at the same phase (ph$=$1.001).

 The sharpness of the fading phase of the He~II intensity curve (Figure 3a) 
 is a challenging characteristic that demands restrictive parameters 
 to be modeled.  For example, if it is 
 due to an eclipse, to be caused by the primary star or by its wind, this 
 wind cannot be isotropic with a mass loss rate of
  $\dot M_{1}$$=$2.5$\times$10$^{-4}$ M$_{\odot}$ yr$^{-1}$
 (PC); this would be optically thick for electron scattering at too large a 
 radius. As an alternative, one could suppose that the Homunculus configuration,
  with a bipolar and equatorial disk could be scaled down to the size of the 
  binary system. In this situation, the polar wind would be responsible 
  (partially of totally) for the eclipse of the He~II emission and the
   equatorial disk, for the X--ray emitting shock wave. Such a configuration 
   has also been proposed on the basis of other observational 
   \citep{S03, vB03} and theoretical \citep{MD01} considerations.
    However, results from spectral modeling in the optical 
   \citep{HDIG} and infrared interferometry \citep{vB03}
   have found mass-loss rates of
  $\dot M_{1}$$\sim$10$^{-3}$ M$_{\odot}$ yr$^{-1}$. Such determinations
  are clearly  in contradiction with our eclipse hypothesis, unless 
  the mass-loss rate is highly anisotropic.

In the wind--wind collision model, the X--ray luminosity is inversely proportional 
 to the separation of the stars, $L_{X}~\alpha~D^{-1}$, so that at periastron, 
 X--ray luminosity is maximum \citep{u92}.  At the same time, the angle subtended 
 by the secondary star as seen from each point of the shock wave is 
 $\alpha~D^{-1}$, making the fraction of the UV and X--ray photons that 
 are absorbed by the wind from the secondary star 
 to go with the square of the inverse of the distance between the two stars. 
 Therefore one expects that the luminosity of the He~II line to be inversely 
 proportional to the cube of the stellar separation: $L_{HeII}~\alpha~D^{-3}$. 
 Figure 3 shows that the rising phase and the peak of the He~II EW curve can be well 
 described by a function of the form $L_{HeII}~\alpha~D^{-3}$ for ecc$=$0.82--0.84.
This agreement, however, does 
 not prove that the He~II emission comes from the wind of the secondary star;
 any model that provides a dependence as the inverse of the cube of the distance 
 would be as good as well.

 In this simplified model we suppose that the emission is highly concentrated. We 
 should keep in mind that there are indications for extended emission as well, 
 so that the real
  picture is actually more complicated, but its detailed analysis and 
  modeling are beyond the scope of this Letter.

\acknowledgments
 We thank FAPESP and CNPq for financial support and to an anonymous referee for 
 useful suggestions.

\end{document}